# INCONSISTENCIES AND ERRORS IN TRADITIONAL APPROACHES TO ENERGY IN OUR INTRODUCTORY COURSES

## Inconsistencias y errores en los enfoques tradicionales sobre la energía en nuestros cursos introductorios

*Álvaro Suárez*
ORCID https://orcid.org/0000-0002-5345-5565
Consejo de Formación en Educación, Instituto de Profesores Artigas, Montevideo, Uruguay
alsua@outlook.com

*Daniel Baccino*
ORCID https://orcid.org/0000-0001-5572-2623
Consejo de Formación en Educación, Instituto de Profesores Artigas, Montevideo, Uruguay
dbaccisi@gmail.com

*Arturo C. Martí*
ORCID https://orcid.org/0000-0003-2023-8676
Instituto de Física, Facultad de Ciencias, Udelar, Montevideo, Uruguay
marti@fisica.edu.uy

*Martín Monteiro*
ORCID https://orcid.org/0000-0001-9472-2116
Universidad ORT, Montevideo, Uruguay
monteiro@ort.edu.uy

**Abstract:**

We present a critical analysis of the classical approaches to energy subjects, based on the work-energy theorem and the conservation of mechanical energy that are proposed in the courses of the first years of tertiary education. We show how these approaches present a series of inconsistencies and errors that are a source of conceptual difficulties in students. We then analyze a modern treatment aimed at mechanical courses based on the results of research in Physics education of the last 40 years. We put special emphasis on the principle of energy conservation as one of the fundamental principles of nature, prioritizing the concepts of system, environment, transfer and transformation of energy.

**Keywords:** Energy, mechanical energy, teaching, energy conservation

**Resumen:**

Presentamos un análisis crítico de los enfoques clásicos de los contenidos de energía, basado en el teorema trabajo-energía y la conservación de la energía mecánica que se proponen en los cursos de los primeros años de educación terciaria. Mostramos cómo estos enfoques presentan una serie de inconsistencias y errores que son fuente de dificultades conceptuales en los estudiantes. Analizamos luego un tratamiento moderno dirigido a cursos de mecánica basado en los resultados de las investigaciones en educación de la Física de los últimos 40 años. Hacemos especial énfasis en el principio de conservación de la energía como uno de los principios fundamentales de la naturaleza, jerarquizando los conceptos de sistema, ambiente, transferencia y transformación de energía.

## 1. The classic approach to energy and its inconsistencies

For several decades, some physicists have led a crusade to expose the shortcomings in the classical approaches with which many of us learned the work-energy theorem and the law of conservation of mechanical energy. Despite a substantial number of articles published in that line (see, for example, Arons, 1989 and 1999; Sherwood, 1983; Sherwood and Bernard, 1984; Jewett, 2008 a, 2008 b; Hetch, 2019; Chabay, Sherwood and Titus, 2019), it took several years before there was a significant impact on the most widely used physics textbooks in basic tertiary-level courses. In recent years, however, more modern approaches based on Physics Education Research (P.E.R.) have gradually emerged.

In our country, Uruguay, the only work published so far revealing some of the shortcomings in the classical approaches is the article "Comentarios sobre el trabajo de las fuerzas aplicadas sobre sólidos reales" (Comments on the work of applied forces on real solids; Núñez, 2011), which appeared in the journal *Educación en Física* of the Uruguayan Association of Physics Teachers (APFU). The article shows the need for using the model of a deformable solid body to explain the existence of the dry friction force and the characteristics of the work it does. These facts, which can be easily understood from the principle of conservation of energy, have profound implications for the way we approach our mechanics courses. Consider, for instance, the process of warming our hands. When we rub them together, if we take one hand as a system, the internal energy of that hand increases as its temperature rises. This increase in energy arises simply from the transfer of energy from the work done by the force exerted by the other hand. Within this simple example, however, lies a problem that we do not generally address until we deal with the first principle of thermodynamics: the only model of a body that allows us to consider changes in temperature is that of a deformable solid, according to which bodies can change their internal energy. This means that the behavior of something as "simple" as a block sliding on a surface cannot be studied energetically without considering the object as a deformable solid body. But what are the implications of this observation?

Newton's laws are at the heart of traditional approaches to mechanics courses. They are the starting point for studying the behavior of bodies, from the simplest to the most complex. Usually, one starts by positing the particle model to deduce the theorems of conservation of mechanical energy and momentum. The rigid body model is then introduced, followed by the study of fluids and thermodynamics, usually relegating the analysis of deformable solids to the far ends of the curriculum. In the latter—which is necessary for modelling daily situations, as shown below—there appear some problems with energy propositions stemming from Newton's laws that lead to inconsistent results. In addition to this problem, as we will discuss later, both the usual definition of the work done by a force and the deduction of the work-energy theorem present limitations and often lead to confusion among students.

In this work we present a modern approach to addressing the concept of energy in basic physics courses at the tertiary level. To that end, the article is structured as follows: in section 2 we present some qualitative examples of the inconsistencies derived from applying the definition of work and the work-energy theorem to deformable solid bodies; in section 3 we introduce the concept of pseudowork and the energy equation of the center of mass; in section 4 we define the energy principle, analyzing the role of the systems and the role of external and internal forces at the core of the principle of conservation of energy; in section 5 we study some classical examples of mechanics, comparing classical and modern approaches; in section 6 we briefly review the historical evolution of the approach to energy, concluding with our final considerations in section 7.

## 2. Some qualitative examples

Imagine that a person on skates who is initially at rest exerts a force $\vec{F}$ on a wall, as shown in figure 1. As a result, the person is driven in the opposite direction, increasing his or her speed and thus his or her translational kinetic energy. If we apply the work-energy theorem to this situation, we will immediately arrive at a contradiction. Since the point of application does not move, the force does not do work. Therefore, although the person increases his or her translational kinetic energy, the network done on him or her is zero.

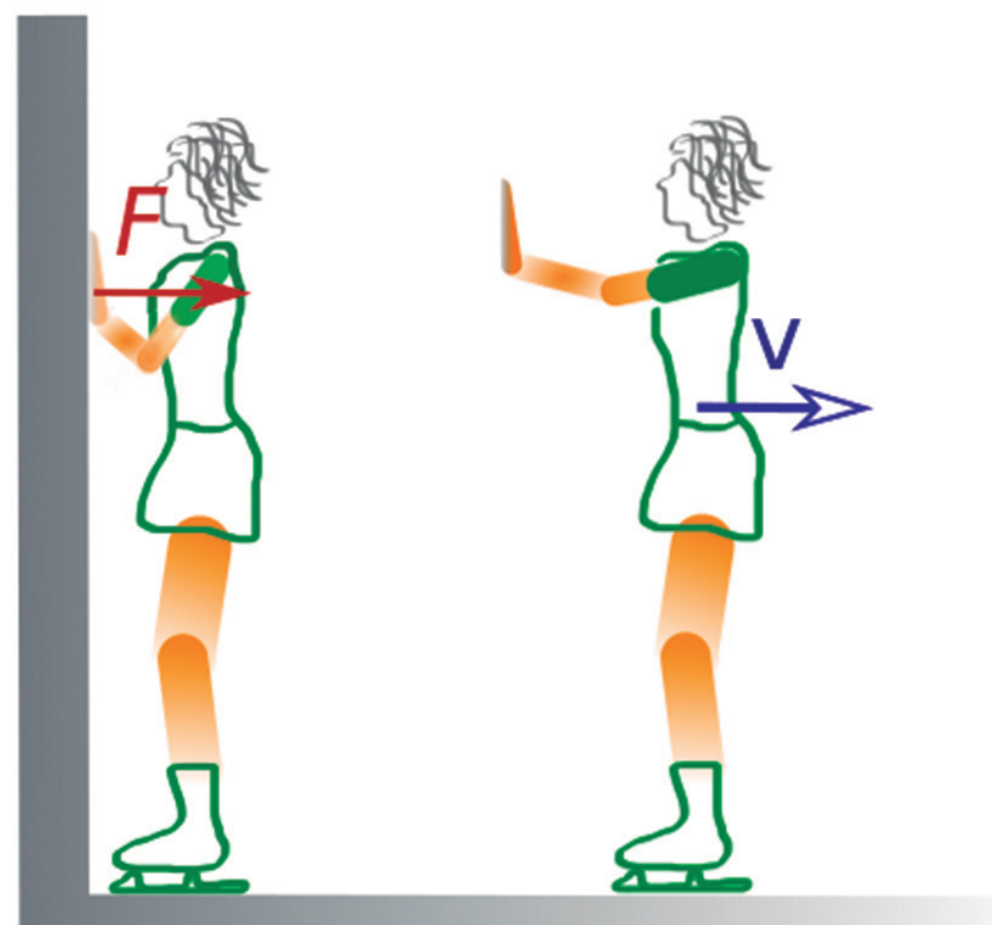

Figure 1. Diagram of a person on skates who moves by exerting a force $\vec{F}$ on the wall.

The above example reveals the inconsistencies associated with the definition of work based on the particle model. In the particle model, the displacement of the point of force application always coincides with the displacement of the particle, whereas in a deformable solid body that is not always the case. This approach to energy, which does not take into account the mechanical aspects of deformable solids, is a source of confusion for students (Lindsey, Heron and Shaffer, 2009; Gutierrez-Berraondo, Goikoetxea, Guisasola and Zavala, 2017). It is also worth noting that the work-energy theorem, in its simplest formulation, is only valid for particles and applies to a limited set of situations.

As a second example, let us consider a block placed on a rugged horizontal surface and pushed by a horizontal force $\vec{T}$ so that it moves along a distance ($d$) with constant velocity, as shown in figure 2 (Resnick and Halliday, 2002).

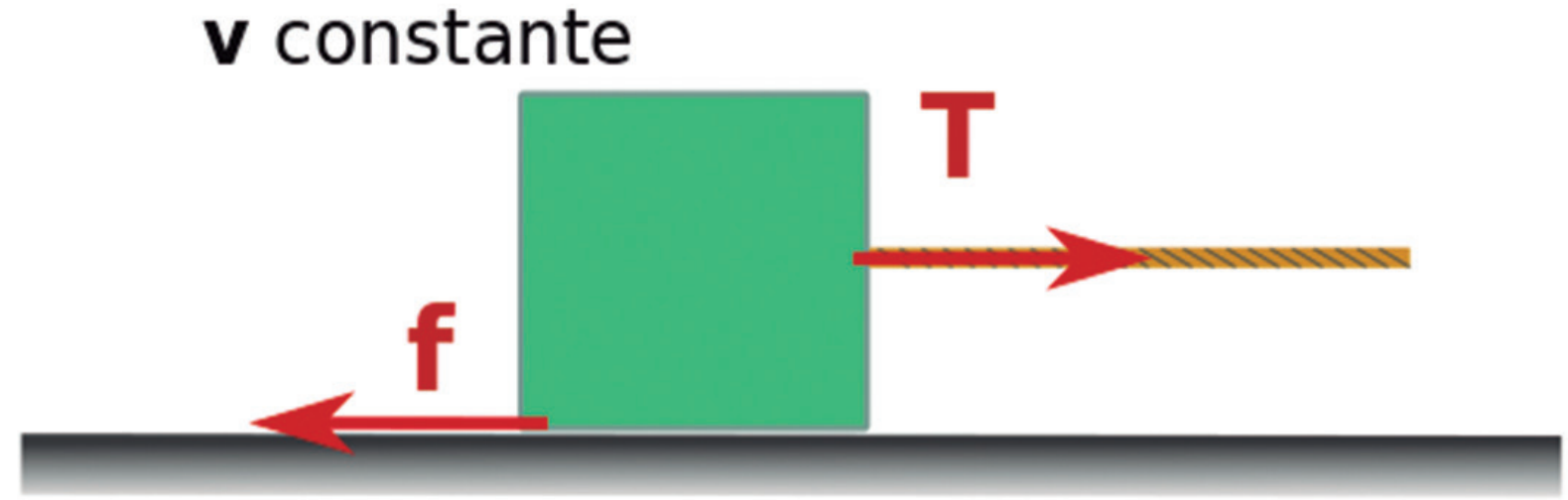

Figure 2. Diagram of a block pulled by a rope and moving with constant speed along a rugged horizontal surface.

If we consider the block as a system and apply the work-energy theorem, we can conclude that the network is zero. We could have deduced this in two ways: on the one hand, by seeing that the speed of the block is constant and, therefore, so is its kinetic energy; or, alternatively, by recognizing that if the block moves at a constant speed, the net force is zero and thus the network is also zero.

Although the previous analysis seems reasonable, it is highly inconsistent from a physical point of view. If the network on the block were zero, there would be no transfer of energy from the surrounding and the total energy of the block would have to remain constant; however, we know that as the block slides on the surface, its temperature increases and, consequently, so does its internal energy. Therefore, we conclude that there should be a transfer of energy as work towards the block, i.e., the *network should be positive*.

Several relevant observations can be made about the previous example. Firstly, it is clear that the work-energy theorem cannot be applied to the situation described, since it is only valid for particles, and the block, given the increase of its internal energy, must be modeled as a deformable solid. Secondly, if the network done on the block is positive, the work of the force $\vec{T}$ must be greater than the work of the friction force. As an immediate consequence of this, the absolute value of the work done by the friction force has to be less than $f_{fric} d$. As a matter of fact, *it is not possible to calculate the friction work exactly, because the effective distance is generally less than the displacement of the block* $f_{fric} d$ (Sherwood and Bernard, 1984). The difference between the usual and the correct calculation could be regarded, beforehand, as rather insignificant. However, it can be shown that the absolute value of the work of the friction force, depending on the characteristics of the two surfaces in contact, can vary between 0 and  (Sherwood and Bernard, 1984). Under certain conditions of symmetry, particularly when two identical bodies slide against each other, it can be easily shown that the work of friction is half the work that would be obtained by classical calculation (Sherwood and Bernard, 1984; Chabay and Sherwood, 2015, p. 391).

As a third and final observation, since the network on the block is positive, the work done by the net force is not equal to the sum of the work done by each force, i.e., even if the net force on the block is zero, the network is not. Although this may seem anti-intuitive, it is directly related to the fact that the equivalence is only valid for particles or rigid bodies, but not for deformable solids.

From the examples presented, it is possible to see the inconsistencies and possible confusion that may arise from applying the work-energy theorem to objects that change their internal energy. The only rightful way to understand these processes is to approach them based on the principle of conservation of energy as explained by the first principle of thermodynamics. This is a general principle that cannot be deduced from Newton's laws. Although in the present work we only present two situations, similar contradictions appear in the energy analysis of most daily life phenomena (e.g., a person walking or jumping, the wheels of an accelerating car spinning without sliding, or simply a ball hitting the ground), when the principle of conservation of energy is not taken as the starting point for the analysis (Güémez, 2013).

Another relevant aspect to be highlighted along with the above limitations is that the classical approaches to energy, based on Newton's laws, generally relegate the concept of system to background, as well as the role of external and internal forces in the transfer and transformation of energy. Familiarity with these concepts is crucial to fully understand the conservation of energy as a fundamental principle of nature.

## 3. Pseudowork and energy equation of the center of mass

A widespread way of presenting the concept of energy in introductory high school and university physics courses is to define the work of a force and demonstrate the work-energy theorem. The work that a force $\vec{F}$ does on a particle, along a trajectory C, is defined as the path integral: $W_F = \int_C \vec{F} \cdot d\vec{r}$. Once this new physical magnitude is established, Newton's second law $\vec{F}_{net} = m\vec{a}$, is applied to develop the work integral of the net force and to arrive at the conclusion that the network is equal to the variation of a certain quantity K, which is equal to ($\frac{1}{2}\, mv^2$). This quantity is called kinetic energy of the particle, with which the work-energy theorem in its first form says that $\Delta K = W_{net}$. The member on the left is a kinematic property of the particle, while the one on the right refers to the dynamics of the interactions between the surrounding and the particle. The work-energy theorem, as presented above, is an integral version of Newton's second law, and as such is universal within the framework of classical mechanics. Although it does not introduce any new physical principle, it proves to be a useful tool in the treatment of numerous situations.

The fact that the work-energy theorem is an essentially dynamic equation may be more evident if we deduce it, for example, from the three Cartesian components of Newton's second law,

$$m\frac{dv_x}{dt}=F_{neta-x} \quad (1)$$

$$m\frac{dv_y}{dt}=F_{neta-y} \quad (2)$$

$$m\frac{dv_z}{dt}=F_{neta-z} \quad (3)$$

By integrating both sides of the equality, we arrive at the following:

$$\frac{1}{2}m\Delta(v_x^2)=\int_C F_{neta-x}\,dx \quad (4)$$

$$\frac{1}{2}m\Delta(v_y^2)=\int_C F_{neta-y}\,dy \quad (5)$$

$$\frac{1}{2}m\Delta(v_z^2)=\int_C F_{neta-z}\,dz \quad (6)$$

by adding the members of the three equations,

$$\frac{1}{2}m\Delta(v_x^2+v_y^2+v_z^2)=\int_C F_{neta-x}\,dx+\int_C F_{neta-y}\,dy+\int_C F_{neta-z}\,dz \quad (7)$$

we arrive at the definitive form of the theorem

$$\Delta K=W_{net} \quad (8)$$

The fact that equation 8 can be expressed in components (equations 4 to 6) makes it clear that, despite its name, the work-energy theorem is a dynamic equation but not a complete expression of the conservation of energy, as mentioned above.

When equation 8, which is valid for a particle, is applied to systems with an internal structure, certain contradictions appear, such as those mentioned in the examples of section 2. We could say that there are at least two ways of avoiding these contradictions: resorting to the correct application of dynamics or to the principle of conservation of energy.

A correct application of dynamics would be to apply Newton's second law for the center of mass of a particle system, $\vec{F}_{net-ext}= m\vec{a}_{CM}$ . With a progression similar to the one followed for a particle, we arrive at an energy equation of the centre of mass,

$$\Delta K_{CM}=\int_C \vec{F}_{neta-ext}\cdot d\vec{r}_{CM} \quad (9)$$

where the left member is the kinetic energy variation of the center of mass and the right member is the path integral of the external net force as if it were applied at the center of mass. This leads to an equation very similar to that of the work-energy theorem, but with an integral that despite being very similar to work should not be confused with it, since the forces are generally not applied at the center of mass. For this reason, many authors refer to that integral as pseudowork (Penchina, 1978; Sherwood, 1983; Arons, 1989; Mallinckrodt and Leff, 1992; Jewett, 2008 a). This highlights an aspect that should always be considered: the fact that the energy equation of the center of mass is an essentially dynamic equation (as is the work-energy theorem) and not a true energy equation.

As a dynamic equation, it is not surprising that the energy equation of the center of mass can be applied without any contradiction to systems with internal structure, such as the examples in section 2. In the case of the skater, the external force exerted by the wall does pseudowork because the center of mass is displaced, and that pseudowork is equal to the kinetic energy acquired by the skater. In the case of the block, the constancy of the kinetic energy results in the fact that the pseudowork done by the net force is zero, from which it is concluded that the friction force has the same modulus as the force $\vec{T}$. These two examples show, once again, that the work-energy theorem and its variant for the center of mass are in fact dynamic equations.

## 4. Principle of energy and the importance of the system's definition

The principle of conservation of energy was formulated based on the works of Joule and other scientists, being valid not only for particles but also for systems with an internal structure. The internal energy is thus introduced, consisting of the sum of the kinetic and potential energies of the microscopic constituents of the system, so that the total energy of the system can be expressed as $E_{system} = K + U + E_{int}$ and its change can be due to multiple forms of energy transfer including work, heat, mass, waves, and radiation, among others (Serway and Jewett, 2015). That is to say:

$$\Delta (K + U + E_{int}) = W + Q + \ldots \quad (10)$$

This expression is called the principle of conservation of energy, or more simply, the principle of energy (Fred Reif, quoted by Sherwood, 2019). Thus, the total energy of a system may not be constant, but its increase or decrease can always be explained in terms of some input or output across the system's boundary. This is the meaning of conservation of energy. *The total energy of a system may not be constant, yet the energy is always conserved.*

From the principle of energy, we can infer, following a different path than usual, the theorem of conservation of mechanical energy. To do this, let us start from equation 10 and consider an isolated system, i.e., a system in which the network of external forces, as well as other forms of energy transfer from the surrounding, are zero:

$$\Delta E_{sistema} = \Delta K + \Delta U + \Delta E_{int} = 0 \quad (11)$$

Being isolated, the only forces that can act between the different parts of the system are internal. These forces, unlike the external ones that can transfer energy to the system by means of work, are the ones that cause the transformations of energy within the system by means of internal work, and can be classified as conservative and nonconservative (Resnick, Halliday and Krane, 2002; Knight, 2004; Jewett, 2008 b).

Let us now assume that the internal forces acting on the system are all conservative. In that case, there are no changes of internal energy, $\Delta E_{int} = 0$, and all the internal work is due to the conservative forces, being therefore associated to the variations in the potential energies of the system. Thus, the only possible energy transformations are from potential to kinetic and vice versa. Equation 11 now becomes:

$$\Delta K + \Delta U = 0 \quad (12)$$

As mechanical energy is the sum of kinetics and potential, we obtain the expression for the theorem of conservation of mechanical energy:

$$\Delta E_{mec} = 0 \quad (13)$$

From the approach presented above, the assumptions on which this theorem is based are evident. It is *only valid for an isolated system where the internal forces at work are all conservative*. Therefore, the theorem of conservation of mechanical energy is clearly not a fundamental principle of physics, but it is a theorem derived from the principle of energy, whose scope of validity is limited.

From a semantic point of view, "conservation of mechanical energy" may lead to conceptual errors, since it is a quantity that, under certain conditions, may or may not remain constant. Therefore, we emphasize that it is not correct to identify a constant magnitude with a general principle of conservation.

We can take the principle of energy, once again, as a starting point, but we assume that the only transfers of energy into the system come from external work, i.e.:

$$\Delta (K + U + E_{int}) = W \quad (14)$$

When we apply this equation, the concrete expression of the different terms depends on the choice of the system boundary. Of course, that boundary is artificial, since we are free to choose the elements that comprise the system, but it is *mandatory* that we make such a choice. Because depending on the system some contributions will remain on the left side of the equation, constituting part of the energy state of the system, and other contributions will remain on the right side, as part of the interactions with the surrounding, which make the system gain or lose energy. Not choosing the system explicitly can lead to errors, as shown in the following example.

Consider the following situation: an apple of mass $m$ that falls to the ground from a height $h$. If the system is only the apple, then there is no change of potential energy or internal energy, and there is an external force which is the one exerted by the Earth, then,

$$\Delta K = W \quad (15)$$

or alternatively

$$\tfrac{1}{2}\, mv^2 - 0 = mgh \quad (16)$$

Conversely, if we consider the apple-Earth system, then there is no external force, although there is a potential energy shift,

$$\Delta K + \Delta U = 0 \quad (17)$$

which easily translates into.

$$\tfrac{1}{2}\, mv^2 - mgh = 0 \quad (18)$$

Evidently, both approaches are valid and lead to the same result, i.e., a final speed is $v = \sqrt{2gh}$

When due attention is not paid to the definition of the system, when the boundary is diffuse or simply not defined, errors such as the following may occur,

$$\Delta K + \Delta U = W \quad (19)$$

i.e., including the interaction of the system as potential energy and, at the same time, as work, which would erroneously result in $v = \sqrt{4gh}$.

## 5. Analysis of a "classic" example

As an example, in this section we analyze, from a near quantitative approach, a situation that usually appears in many textbooks. Based on the propositions discussed in the previous sections, we study, from an energy point of view, some characteristics of the movement of a child sliding down a slide, shown schematically in figure 3.

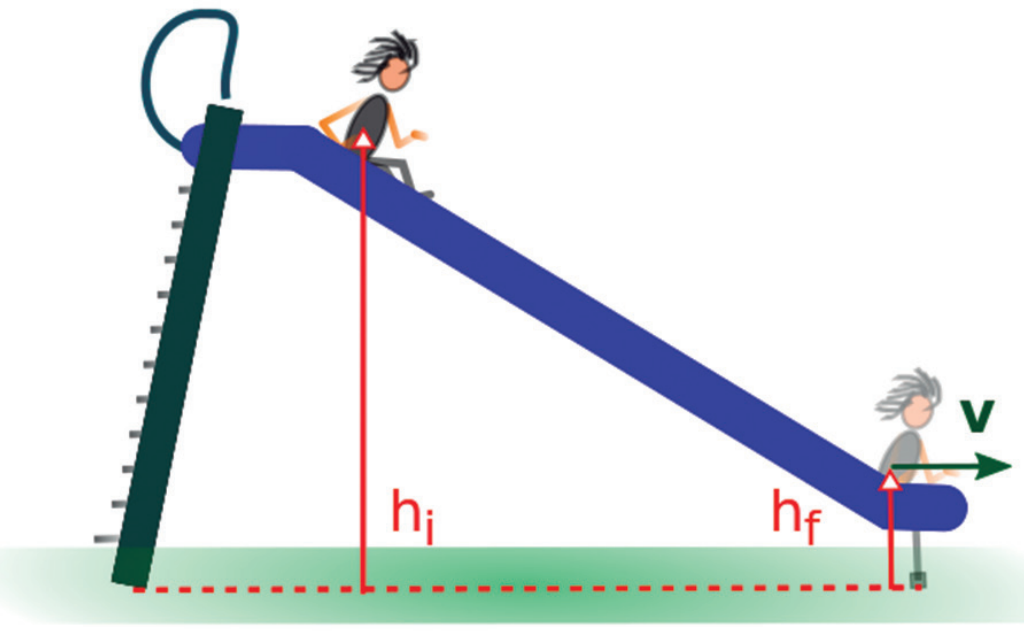

Figure 3. Diagram of a child sliding down a slide.

The analysis takes the principle of energy presented in section 3 as the starting point. We then introduce some simplifying hypotheses to problematize the situation, depending on how we define the system and its respective surrounding. We show some limitations, advantages and disadvantages, in terms of that essential initial definition, which should also be made explicit in this energy context.

There is at least one other aspect that we should consider carefully. The mechanical energy dissipated due to the existence of kinetic friction between the child and the slide is transformed into internal energy *for both bodies*, modelled as deformable solids. If it were analyzed according to an introductory course, it would not be easy to know which part of that energy goes to the child and which part is transferred to the slide.

From now on we will assume some simplifying hypothesis that will help us to highlight some aspects of the situation. We will accept as valid that the effect of air on the child is irrelevant. Furthermore, we postulate that the energy transfers between the defined system and its surrounding are carried out exclusively through work (for example, that the temperatures of both bodies are similar, so there is no transfer of energy through heat). The latter assumption may be questionable, especially for some defined systems, but we make it in order to highlight the aspects that are relevant in the context of the article.

We will now analyze the proposed situation under different choices of system.

**5a. Child system and Child Earth system**

If we define the system "child", its surrounding comprises the slide and the Earth. In this case, there are two works that cross the system's boundary. They are, therefore, external works. The gravitational work can be determined if we know the weight of the child and the displacement of its center of mass. Reaffirming the example of the apple at the end of section 4, no gravitational potential energy is defined for this system.

As we saw in section 2, there are significant difficulties in determining the work associated with the force of kinetic friction. If we consider the usual approach of asking what is the speed of the child at the bottom of the slide, we can use the principle of energy as a starting point and add the aforementioned definitions and hypotheses to conclude that it is not possible to determine said speed (at least exactly) because the energy transferred to the system by the effects of friction is unknown.

For the definition of the "child-Earth" system, its surrounding is exclusively the slide. Therefore, to apply the *principle of energy* we must know the work done by the force of friction made by the slide. We find here the same difficulties referred to in the previous paragraph.

**5b "Child-Slide-Earth" system**

This choice, together with the hypotheses we define, implies that there is no transfer of energy between the system and its surrounding, i.e., it is an *isolated system*. This choice does not require the determination of the work of friction, the difficulty of which we have already noted.

Regarding the energy transformations inside the system, changes in the kinetic energy, in the gravitational potential energy, and in the internal energy take place. The starting point to propose a solution to the problem is the *principle of energy*, in this case for an isolated system: $\Delta E_{system} = 0$. Given the assumptions made when presenting the problem, the *child-slide system is the only one that keeps the energy constant*.

How do we formulate the transformation of energy in the system? If we compare an initial situation (the child at the top of the slide) with a final situation (the child coming out of the slide), we can identify a change in the gravitational configuration, therefore $\Delta U_g \neq 0$. The kinetic energy $K$ of the system also changes if we measure it from a fixed reference to the floor. Finally, there is an increase in the internal energy of the system due to the effect of friction on the sliding bodies, associated with temperature increases of the child and the slide.

In our elementary model, the *principle of energy* is expressed as:

$$\Delta U_g + \Delta K + \Delta E_{int} = 0 \quad (20)$$

The first term has a negative sign since the parts of the system move closer; it can be expressed as $mg\,(h_f - h_i)$. The second term has a positive sign because the child travels faster at the end than at the beginning; if the child starts moving from a state of rest, the term is $\frac{1}{2}\,mv^2$[1]. Particular attention should be given to the third term (also positive), which corresponds to the variation of the child's internal energy and the slide. It can be demonstrated, based on Newton's second law and other energy considerations (Tipler and Mosca, 2010; Serway and Jewett, 2015; Chabay and Sherwood, 2015; Knight, 2017), that this variation is given by

$$\Delta E_{int} = f_{fric}\, d \quad (21)$$

being $d$ the distance traveled by the child on the slide.

Note that the product $f_{fric}\, d$ is not equal to the absolute value of the work of the friction force exerted by the slide on the child; it is the variation of the internal energy of the two bodies that interact through the friction force. It is clear then (as mentioned in section 2) that taking the product $f_{fric}\, d$ as the absolute value of the work of the friction force exerted by the slide on the child is physically incorrect.

Having defined the energy variations, we are able to formulate equation 20 in terms of quantities that are usually already known in a "sample" problem:

$$mg\,(h_f - h_i) + \frac{1}{2}mv^2 + f_{fric}\, d = 0 \quad (22)$$

1 Assuming that the child's kinetic energy can be properly expressed by a translational term exclusively.

This approach makes it possible, for example, to determine the child's final speed if the rest of the quantities are known:

$$v = \sqrt{\frac{2}{m}\left(-\left(mg\left(h_f - h_i\right) + f_{roz}\, d\right)\right)} \quad (23)$$

If we consider exclusively the final expression, we see that it is the same as that which would be obtained with other approaches, such as that in which the system has not been defined in the same way, or that in which the child has been modelled as a particle, or that in which $\Delta E_m = W_{nc}$. However, the fact that we arrive at a correct result does not mean that the process leading to it is also correct. In this sense, we want to emphasize that the classical approach that leads to equation 23, based on the fact that the work of friction, is given by $-f_{fric}\, d$ is not correct. The quantity $f_{fric}\, d$ has a very different physical meaning from the one it is usually given, i.e., it is equal to the variation of the internal energy of the two bodies sliding on each other.

From what has been exposed in this work, we see how only an approach based on the principle of energy allows us to root out the inconsistencies from the classical approach to the subject that do not contribute to its conceptual understanding by the students.

In the following section, we will review the evolution of the treatment of this subject, focusing on an example that many readers will find paradigmatic: the first volume of the classic *Resnick-Halliday* textbook.

## 6. Historical evolution

> She said, *"Suppose you push a block across the floor at constant speed. The net force (your push and the opposing friction force) is zero, so choosing the block as the system no work is done, yet the block's temperature rises, so the internal energy is increasing. I'm very confused." I said, "Oh, I can explain this. You just, uh, well, you see, uh.....I have no idea."*

The previous quotes is taken from an anecdote that Bruce Sherwood shares in his blog (Pseudowork and real work, 2017), from when he was responsible for a teaching project. He quotes an exchange dating back to 1971 between an undergraduate student, Lynell Cannell, and Sherwood himself. The student's assignment was to write a tutorial on energy, and her lack of progress on the work led to the exchange. The interest of the project's leader in the subject led Sherwood to write, not without difficulty, two of his best-known articles on these topics (Sherwood, 1983; Sherwood and Bernard, 1984)

In 2017, Sherwood published a post on his personal blog where he transcribes part of a response from D. Halliday to a letter Sherwood sent him in 1983, in which he raised his concerns about the treatment of the subject of energy in deformable systems in classic textbooks, such as the one Halliday had published jointly with Resnick. In his response, Halliday wrote:

> *Let me say at once that we are well aware of its serious flaws, along precisely*

*the lines that you describe. We have tried several times to patch things up in successive printings but the matter runs too 66deep for anything but a total rewrite. We have, in fact, such a rewrite at hand, awaiting a possible next edition.*

Other contributions can be identified in this process of reviewing the way energy is approached in introductory courses; we shall mention three examples. *Energy and the Confused Student* is a series of five articles that refer to the confusion on the subject among students. In our list of references, we cite the first two of the five issues (Jewett, 2008 a and 2008 b), which deal with the concepts of work and systems. The third example is *Developing the energy concepts in introductory physics* (Arons, 1989). This revisionist process, which has been illustrated here by citing some articles, has also permeated the newer editions of some textbooks currently used in introductory physics courses.

A typical example, for several generations of students, is the classic *Halliday-Resnick*[2], where structural changes can be identified when reviewing the editions of its first volume. For an initial approach to the problem, we examined the thematic table contents of the first five editions in Spanish, from the 60s through the 2000s. This overview allowed us to identify two subsets of editions. The first four editions include a similar structure in terms of content. A change of structure in the treatment of the topic of energy is evidenced from the fifth English edition (corresponding to the fourth Spanish edition) (Resnick, Halliday y Krane, 2002).

Which qualitative aspects regarding energy stand out in the first four editions as a whole? The first chapter in which the subject is explained is *Work and Energy* (chapter 7 in all three editions). It can be considered the closing of a cycle focused on the treatment of particle movement, whose development initially includes a kinematic approach followed by a dynamics perspective. The following chapter, *Conservation of Energy*, deals essentially with conservative systems, introducing the concepts of conservative force and potential energy. The concept of internal energy is explained later in chapter related to thermodynamics (Chapter 22: *Heat and the first law of thermodynamics*, in the fifth English edition).

Our general overview of the last editions show at least two aspects that differ from the first editions. The three chapters that initially refer to energy appear farther away from the beginning (chapters 11 to 13 in the fitth English edition), after chapters referring to particle systems (kinematics and rotational dynamics). Chapter 13 develops a novel treatment in comparison with the previous editions, which could be considered a "preview of the first principle of thermodynamics".

The last statement of the previous paragraph requires some explanation. The first section, *Work done on a system by external forces*, conveys the idea that the system in its surrounding is limited by a boundary, and that work is a way to measure the transference. The second section, *Internal energy of a particle system*, extends the approach beyond mechanical energy and into the macroscopic realm, while *Frictional work* is perhaps the chapter's most dissenting section, showing aspects that we have mentioned before in this work (Sherwood and Bernard, 1984; Núñez, 2011), in reference to the inconsistencies associated

2 This is how we refer to the complete set of five editions, detailed in our list of References.

with the previous approach. The section *Energy of the center of mass* presents an approach that we have mentioned in this work (Sherwood, 1983), addressing some questions about deformable systems. The chapter closes with the section *Energy transfer by heat*, in which the first law of thermodynamics is presented. The first law of thermodynamics reappears in a later chapter near the end of the volume.

Perhaps the most relevant conceptual change observed in our general overview of *Halliday-Resnick* is that the last chapter on energy in the 5$^{th}$ English editions is closer to living up to its name than in the first editions. This is because in the first editions the focus is on the conservation of mechanical energy, whereas in the more recent editions the focus is on the idea of the conservation of energy.

In the previous paragraphs we presented a transformation in one of the "classic" texts. This revision of the treatment of the subject of energy can also be seen in other books used at this level, such as Serway and Jewett (2015) or Tipler and Mosca (2010). An example of an introductory text that considers several of the guidelines mentioned in this work, particularly in reference to the treatment of energy, is *Matter and Interactions* (Chabay and Sherwood, 2015).

We close this section with a quote that invites us to reflect on the convenience of revising curricular plans associated with the subject of energy, given the importance that this has in the contribution that science makes to understanding the environment, not only concerning our discipline. According to Chabay and Sherwood (2019),

> *Rather than attempting to add activities that give students practice in dealing with such confusion*[3]*, we advocate a restructuring of the energy component of the introductory physics curriculum in a manner that is coherent, consistent, and contemporary, and which empowers students to analyze interesting phenomena such as fission and fusion simply by applying fundamental principles.*

## 7. Concluding remarks

We began this article by showing the most important inconsistencies of the treatments of energy derived from Newton's laws when applied to deformable solids, and then presented a modern approach that results from Physics Education Research (P.E.R.). From this new perspective, the conservation of mechanical energy is not derived from Newton's laws, but rather from the principle of conservation of energy, emphasizing concepts that have not been traditionally considered, such as system, surrounding, transfer and transformation of energy.

In traditional courses, the principle of conservation of energy is not addressed until the study of thermodynamics is finally introduced. One of the most important principles of nature should certainly play a much more central role in physics courses. In this sense, approaching it after studying point mechanics allows the students to have a more general vision of mechanics and be able to study and understand more complex problems. On the other hand, the principle of conservation of energy, given its nature, enables interdisciplinary dialogue and multidisciplinary approaches, so necessary for a quality scientific training.

3 The authors previously list inconsistencies that may cause some of the learning issues observed among students of this subject.

As a final reflection, we observe that education, contrary to other activities, has always been propelled by sheer inertia, which slows down the processes of change. The teaching of physics has not escaped this. As we discussed in the present work, it took nearly 20 years since the first article pointing out the inconsistencies and errors in the usual treatments of energy was published in the *American Journal of Physics* for the new approaches to appear in some of the main texts on General Physics with which many of us were trained. These changes are not random, but they have been appearing concurrently with the development of P.E.R. that has been taking place worldwide. Many aspects of the classic textbooks are slowly beginning to be restructured and changed based on the results of research. At the same time, new texts with completely new and research-based approaches have appeared, for example, *Matter and Interactions*, in which the treatment of mechanics is focused on the principles of conservation instead of on Newton's laws.

As with any process of change—and now speaking specifically of us, the teachers—it is usually more difficult to assimilate and put into practice new approaches when we have been trained in a different way. First, we must have total conviction about these new approaches and the shortcomings of the previous ones to be able to carry out the process of didactic transposition with confidence. In this sense, we are convinced that an in-depth reflection on the subject addressed in this article, based on the results arising from P.E.R., it is paramount to continue training ourselves, improving in our task and bringing our students closer to a more modern and comprehensive understanding of physics.

**Acknowledgements**

We thank ANII and CFE for the financial support to the project "Knowing and influencing the epistemological conceptions of future physics teachers" (FSED_3_2019_1_157320).